\documentclass[journal,comsoc]{IEEEtran}
\usepackage[T1]{fontenc}

\usepackage{graphicx}
\usepackage{xcolor}
\usepackage{hyperref}
\usepackage[utf8]{inputenc}
\usepackage{subcaption}
\usepackage[printonlyused]{acronym}

\ifCLASSINFOpdf

\else

\fi

\usepackage{amsmath}
\usepackage[cmintegrals]{newtxmath}

\acrodef{SDN}{Software Defined Networking}
\acrodef{NFV}{Network Function Virtualisation}
\acrodef{ISP}{Internet Service Provider}
\acrodef{CI/CD}{Continuous Integration/Continuous Deployment}
\acrodef{NGCO}{Next Generation Central Office}
\acrodef{VCO}{Virtual Central Office}
\acrodef{TR}{Technical Report}
\acrodef{cloud-CO}{Cloud Central Office}
\acrodef{ROADM}{Reconfigurable Add Drop Multiplexer}
\acrodef{ForCES}{Forwarding and Control Element Separation}
\acrodef{RCP}{Routing Control Platform}
\acrodef{PCE}{Path Computation Element}
\acrodef{API}{Application Programming Interface}
\acrodef{VNF}{Virtual Network Function}
\acrodef{C-RAN}{cloud Radio Access Networks}
\acrodef{RAN}{Radio Access Networks}
\acrodef{CORD}{Central Office Re-architected as a Data Centre}
\acrodef{BBF}{Broad Band Forum}
\acrodef{DSLAM}{Digital Subscriber Line Access Multiplexer}
\acrodef{OLT}{Optical Line Terminal}
\acrodef{PON}{Passive Optical Network}
\acrodef{MPLS}{Multi Protocol Label Switching}
\acrodef{BNG}{Broadband Network Gateway}
\acrodef{TIP}{Telecom INFRA project}
\acrodef{OLS}{Open Line System}
\acrodef{DPDK}{Data Plane Development Kit}
\acrodef{DMA}{Direct Memory Access}
\acrodef{ONF}{Open Networking Foundation}
\acrodef{NETCONF}{Network Configuration Protocol}
\acrodef{BGP}{Border Gateway Protocol}
\acrodef{SNMP}{Simple Network Management Protocol}
\acrodef{TL1}{Transaction Language 1}
\acrodef{CLI}{Command Line Interface}
\acrodef{OSM}{Open Source MANO}
\acrodef{MANO}{Management & Orchestration}
\acrodef{VNFM}{Virtual Network Function Manager}
\acrodef{NFVO}{Network Function Virtualisation Orchestrator}
\acrodef{VIM}{Virtualized Infrastructure Manager}
\acrodef{R-CORD}{Residantial CORD}
\acrodef{VOLTHA}{Virtual OLT Hardware Abstraction}
\acrodef{M-CORD}{Mobile CORD}
\acrodef{E-CORD}{Enterprise CORD}
\acrodef{VPN}{Virtual Private Network}
\acrodef{SD-WAN}{Software Defined Wide Area Network}
\acrodef{RRU}{Remote Radio Unit}
\acrodef{OPNFV}{Open Platform for NFV}
\acrodef{ONAP}{Open Network Automation Platform}
\acrodef{OPEN-O}{Open Orchestrator}
\acrodef{BBU}{Base Band Unit}
\acrodef{OSNR}{Optical Signal to Noise Ratio}
\acrodef{NGMN}{Next Generation Mobile Network forum}
\acrodef{ODTN}{Open and Disaggregated Transport Network}
\acrodef{QoS}{Quality of Service}
\acrodef{DC}{Data Centre}
\acrodef{DBA}{Dynamic Bandwidth Allocation}
\acrodef{vDBA}{virtual DBA}
\acrodef{DU}{Distributed Unit}
\acrodef{CU}{Centralised Unit}
\acrodef{WSS}{Wavelength Selective Switch}
\acrodef{WEF}{World Economic Forum}
\acrodef{IT}{Information Technology}
\acrodef{MAC}{Medium Access Control}
\acrodef{CO}{Central Office}
\acrodef{QoT}{Quality of Transmission}
\acrodef{DC}{Data Centre}
\acrodef{ODN}{Optical Distribution Network}
\acrodef{IoT}{Internet of Things}
\acrodef{TAPI}{Transport API}
\acrodef{P2P}{Point-to-Point}
\acrodef{SRI-OV}{Single Root Input/Output Virtualisation}
\acrodef{NUMA}{Non-uniform memory access}
\acrodef{VNO}{Virtual Network Operator}
\acrodef{FPGA}{Field Programmable Gate Array}
\acrodef{XGEM}{XG-PON encapsulation method}
\acrodef{XGTC}{XG-PON traffic container}
\acrodef{ONU}{Optical Network Unit}
\acrodef{BMap}{Bandwidth Map}
\acrodef{DBRU}{Dynamic Bandwidth Reporting Unit}
\acrodef{ME}{Merging Engine}
\acrodef{I/O}{Input/Output}
\acrodef{TCont}{Traffic Container}
\acrodef{TDM}{Time Division Multiplexing}
\acrodef{GPP}{General Purpose Processor}
\acrodef{AWS}{Amazon Web Services}
\acrodef{vCD}{vCloud Director}
\acrodef{ECOMP}{Enhanced Control, Orchestration, Management \& Policy}
\acrodef{TCD}{Trinity College Dublin}
\acrodef{T-CONT}{Transmission Container}
\acrodef{LXC}{Linux Container}

\begin{document}

%
\title{Moving the Network to the Cloud: \\
the Cloud Central Office Revolution and its Implications for the Optical Layer}
%
%
%

\author{Marco~Ruffini,~\IEEEmembership{Senior~Member,~IEEE}, Frank~Slyne,~\IEEEmembership{Member,~IEEE,}
\thanks{Prof. Marco Ruffini and Dr. Frank Slyne are with the CONNECT telecommunications research centre, at the University of Dublin, Trinity College, Ireland. Contact email address: marco.ruffini@tcd.ie, fslyne@tcd.ie}}

%
%

\markboth{IEEE Journal of Lightwave Technology, January~2019}%
{Shell \MakeLowercase{\textit{et al.}}: Bare Demo of IEEEtran.cls for IEEE Communications Society Journals}


\maketitle

\begin{abstract}

\ac{SDN} and \ac{NFV} have recently changed the way we operate networks. By decoupling control and data plane operations and virtualising their components, they have opened up new frontiers towards reducing network ownership costs and improving usability and efficiency. Recently, their applicability has moved towards public telecommunications networks, with concepts such as the \ac{cloud-CO} that have pioneered its use in access and metro networks: an idea that has quickly attracted the interest of network operators. By merging mobile, residential and enterprise services into a common framework, built around commoditised data centre types of architectures, future embodiments of this \ac{CO} virtualisation concept could achieve significant capital and operational cost savings, while providing customised network experience to high-capacity and low-latency future applications.

This tutorial provides an overview of the various frameworks and architectures outlining current network disaggregation trends that are leading to the virtualisation/cloudification of central offices. It also provides insight on the virtualisation of the access-metro network, showcasing new software functionalities like the virtual \ac{DBA} mechanisms for \acp{PON}. In addition, we explore how it can bring together different network technologies to enable convergence of mobile and optical access networks and pave the way for the integration of disaggregated \ac{ROADM} networks. 

Finally, this paper discusses some of the open challenges towards the realisation of networks capable of delivering guaranteed performance, while sharing resources across multiple operators and services. 
     
     
\end{abstract}


%
\IEEEpeerreviewmaketitle

\section{Introduction}
%
%
%
%
\IEEEPARstart{T}{he} \ac{SDN} and \ac{NFV} concepts have changed the way we design and operate networks. While they have only recently gained widespread dissemination, much of the underlying work that led to their success dates over twenty years back. 
If we look at the past, by the late 1990's, the Internet, which  was originally intended as a platform for research had become stagnant and closed to innovation. The critical routers and switches of large telecommunication operators and \ac{ISP}s had become overblown monolithic platforms that  required very specialist skills to configure and diagnose when things went wrong.  Since the Internet was no longer just a research network, but was  supporting critical commercial and government services, there  was  no clear governance over how systemic issues in the technology and processes could be rectified or how deficiencies could be improved. 
Thus, novel design ideas, such as that of programmable networks, started to rise in the late '90s, for example with the idea of active networks \cite{SwitchWare}, \cite{NetScript}, where network nodes would expose their resources and capabilities through programming interfaces, allowing online modifications to routers' behaviour. Shortly after, around the beginning of the 2000s decade, work started to appear on the separation of control and data planes, as network operators were looking for methods to increase their control over the network (e.g., with respect to legacy distributed routing algorithms), for traffic engineering purpose. On one hand, this led to the development of open interfaces between control and data plane, driven by projects like the \ac{ForCES} \cite{ForCES}. On the other hand, we saw the rise of centralised network control systems, including the \ac{RCP} \cite{RCP} and the \ac{PCE} \cite{PCE}. These trends were also supported by new hardware developments at that time: the fast increase in capacity requirements drove vendors to develop packet forwarding logic in hardware, leading to its separation form the software control plane; in addition, the commoditisation of the computing platforms meant that general purpose servers could provide more memory and processing resources than embedded routing processors. This is also the time when vendors of commodity \acp{GPP} ) (i.e., Intel and AMD) included instruction sets in their processors for hardware support of virtualisation, up to the standard defined by Popek and Goldberg \cite{PopekGoldberg}. 
Prior to this, virtualisation required specific mainframe hardware (for instance, running an IBM/370 mainframe), or the customisation of applications and operating systems, a process which is both expensive and difficult to maintain.
For an in depth survey of the road to \ac{SDN} the readers should refer to \cite{Femster-road-to-SDN}.

Building upon this previous work, the introduction of the OpenFlow protocol \cite{Openflow-stanford}, towards the end of the 2000s, started the \ac{SDN} revolution in telecommunications networks. The novelty, with respect to previous attempts and ideas, was to provide \acp{API} that could interface with existing off-the-shelf hardware switches, rather than requiring new hardware. Taking advantage of the trend to build switches using merchant silicon, OpenFlow opened up a new world of possibilities for academic researchers. It provided the ability to develop and test new ideas over real networks, paving the way to many research projects across the globe developing new network protocols and functionalities. Within only a few years, \ac{SDN} managed to grasp the attention of the data centre and telecomms industry. For operators, \ac{SDN} is an enabler for the application of the DevOps principles and \ac{CI/CD} to the management and configuration of their networks. Such principles were already being applied to other systems in their companies, such as databases, software platforms and web sites. The practice of Continuous Integration is where changes and releases of applications are continuously tested against a complete set of functional and unit tests. This facilitates the release of features to the production system or network on daily basis, and offers the ability to regress to a snapshot of a previous release easily. \ac{SDN} can thus enable similar innovation on the network control and management operations.


In the mean time, in parallel with the control plane programmability offered by \ac{SDN}, the concept of data plane programmability and virtualisation also continued to develop. Extending the concept of virtualisation of computing environment, where multiple virtual machines could run independently over the same server, the idea of a network hypervisor (or FlowVisor in OpenFlow terms) was brought in to allow different controllers to operate the same physical switch. When applied to a complex network of several nodes, e.g., in a data centre environment, network virtualisation enabled the creation of an abstraction layer that could be used to setup several virtual networks across shared hardware infrastructure. With the support of appropriate software \cite{NSX}, virtual networks could be dynamically created to connect virtual machines; they could be modified online when the associated virtual machines were migrated, or their number was increased or decreased. Coupled with \ac{SDN}, network virtualisation enabled unprecedented network programmability and flexibility in the data centre environment.

The next step was the move of the \ac{SDN} and virtualisation concepts out of the data centre, to the public network. 
Pressed by progressively squeezed operation margins, the operators had been looking for ways to reduce capital and operational expenditures and for means to generate new revenue in their network. \ac{NFV} \cite{ETSI-NFV} aims at virtualising typical functionalities of telecommunications networks, so that they can be decoupled from the hardware. Since most of today's network functions operate in the digital domain, \ac{NFV} allows moving them from expensive dedicated hardware to commodity servers. In addition, the ability to run functions as software over a shared compute infrastructure, facilitates the creation of new services by dynamic composition of chains of \acp{VNF}. The use cases are several: multi-tenancy, as different software instances can be handed to virtual operators, which can have greater control over their virtual slice; a multi-service platform, as different services can be provided with customised resources to meet their requirements of compute and networking capacity; improving the efficiency of running cloud Radio Access Networks (cloud-RAN), as the location of their protocol stack functions can be optimised depending on actual requirements (e.g., real user demand) and computing/networking resource availability. 

Indeed much work is currently focusing on the development of \ac{NFV} software platforms that promise to deliver a comprehensive infrastructure to handle the requirements of current and future telecommunications and service operators.
The \ac{NGCO} is the generic term for the re-architecture of the Telco Central Office towards a fibre-rich, software-centric \ac{CO} that benefits from the principles of virtualisation that have been developed in the data centre. Today we see many different projects from different organisations developing this idea into well-defined architectures and software systems, such as the \ac{ONF} \ac{CORD} \cite{CORD}, the \ac{OPNFV}'s \ac{VCO} \cite{OPNFV} and the \ac{BBF}'s \ac{cloud-CO} \cite{TR-384}.
\ac{CORD} was one of the first projects to pioneer the use of \ac{NFV} in access and metro networks, bringing it inside the central office. This idea has quickly attracted the interest of the networking industry, gaining in only a few years the support of several operators across the world, many of which have started carrying out network trials. This has also led to novel standardisation activities: for example the \ac{BBF} has recently released the \ac{cloud-CO} 
\ac{TR} \cite{TR-384}. 

 Both \ac{VCO} and \ac{CORD} are open source projects with real code bases, while the \ac{cloud-CO} defines standards, through \ac{TR}s, for interoperability (for instance through YANG schemes) and for how the \ac{CO} should function (for instance, sizing and scalability). Practically, both \ac{VCO} and \ac{CORD} use OpenStack as a virtualisation platform, however from the perspective of controllers, \ac{VCO} uses OpenDayLight while \ac{CORD} uses ONOS.

The concept of central office virtualisation brings together different network technologies, providing functional convergence for mobile and optical access networks and paving the way for its extension towards disaggregated optical networks (e.g., the use of \ac{ROADM} white boxes at the access and metro network). 
However many challenges remain to be addressed in order to guarantee the quality of service required to run upcoming 5G applications, while multiplexing network and processing resources across multiple operators and services.

\vspace{3mm}

As it will be clarified in the next section, where we delve into the architectural details of central office virtualisation, while \ac{SDN} and \ac{NFV} can in principle operate independently, they are highly synergistic in a \ac{cloud-CO} environment. Although their definition is often somewhat arbitrary, \ac{SDN}'s task is typically that of providing a software interface to physical devices and thus enable the centralisation of the control plane across multiple devices. \ac{NFV} on the other hand provides virtualisation of the physical hardware, to represent its functionality in software. \ac{NFV} brings advantages associated to increased flexibility in the data plane, allowing cost reduction through the use of commodity servers and enabling resource slicing, thus assigning different network instances to different tenants and services. \ac{SDN}, in parallel, brings in advantages of control plane flexibility, enabling coordination of slices both in single and multi-domain environments and facilitating multi-tenancy by offering network control to multiple entities through the use of programmatic interfaces.
The meaning of slicing can be summarised by the 3GPPP definition \cite{28_801} of "transforming the network/system from a static one size fits all paradigm, to a new paradigm where logical networks/partitions are created, with appropriate isolation, resources and optimised topology, to serve a particular purpose or service category or even individual customers".  While the 3GPPP definition refers inherently to a mobile system (inclusive of Radio Access Network, Core Network Control Plane and User Plane Network Functions), in this paper the concept of slicing extends to additional technologies (e.g., optical backhaul/metro, layer 2 and layer 3 networks) that are part of the end-to-end service path.


In the reminder of the paper we will use the term \ac{cloud-CO} interchangeably with \ac{VCO} or \ac{NGCO}, to refer to the general concept of virtualisation of the central office. 
In the next section, this tutorial paper will introduce a number of different development frameworks that implement the \ac{cloud-CO} concept. We then extend the disaggregation to the optical layer, briefly mentioning some of its pros and cons and their importance towards the realisation of a fully virtualised network. After delving into some technical details on two use cases for network virtualisation and slicing, we give, in section V, a general overview of the economic benefit that the digitisation of the telecommunications industry could bring. Finally, we conclude the paper by exploring some of the outstanding challenges we believe should be addressed in the near future. 



\section{Cloud Central Office Architectures}
A modern \ac{CO} is a network node that terminates residential and business subscriber lines. It typically contains equipment such as \acp{DSLAM}, used to terminate copper broadband lines; \acp{OLT}, used to provide \ac{PON} or point-to-point optical access; data aggregation equipment, typically operated through Ethernet switching; data transport equipment such as \ac{MPLS} routers, used to provide transport data services within the operator's network. It can also contain \acp{BNG}, which provides subscribers connectivity to the Internet.

A \ac{cloud-CO} is a framework for bringing \ac{NFV} into a telecommunications central office, where functions that typically run on dedicated hardware are moved to software frameworks running on commodity hardware.
\begin{figure*}[t]
   \centering
        \includegraphics[width=160mm]{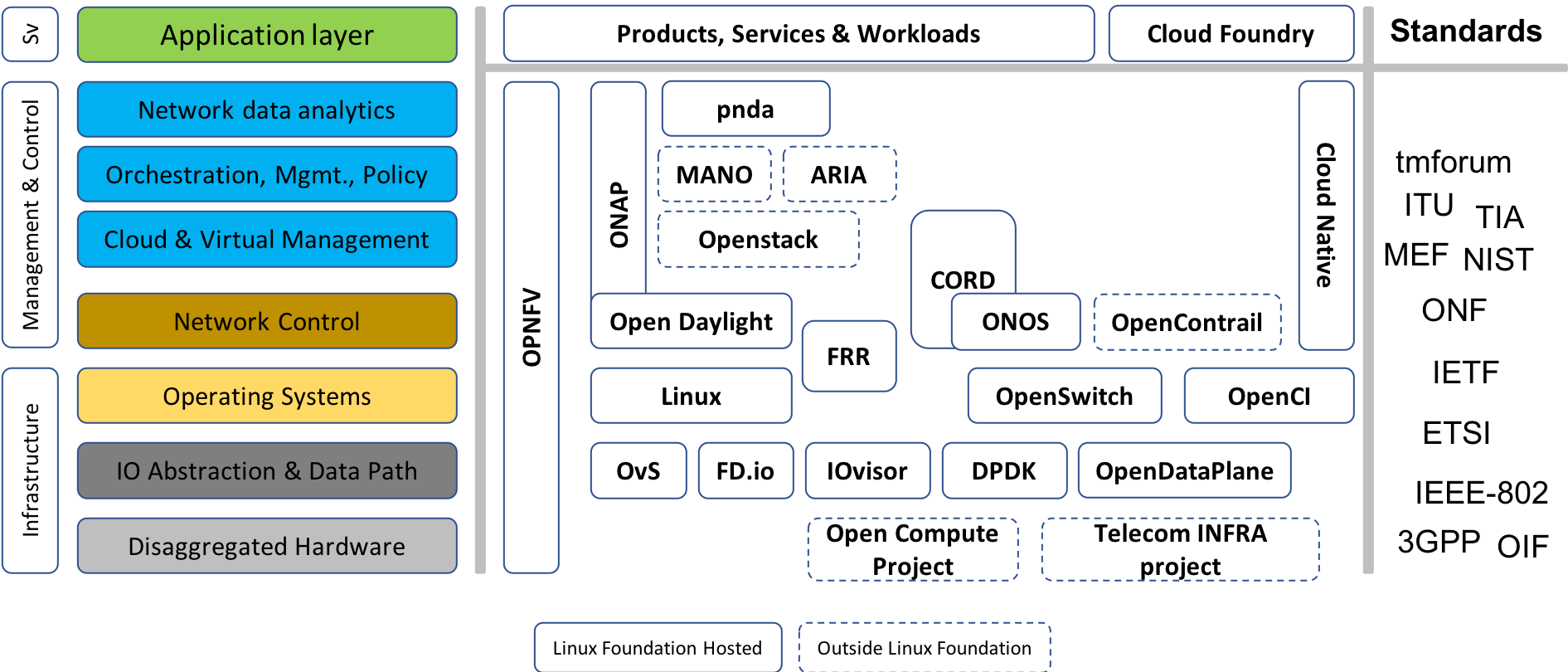}
    \caption{Classification of \ac{NFV}-related development frameworks \cite{Linux_fund_source}}
    \label{fig:Linux_fun_classification}
\end{figure*}
This moves the \ac{CO} architecture towards that of a data centre. Its implementation relies on the development of several software components that closely inter-operate to deliver an end-to-end solution. The diagram in Fig. \ref{fig:Linux_fun_classification} (re-drawn from source \cite{Linux_fund_source}), is an attempt to map the functionality of some of these components with respect to the infrastructure, management \& control plane stack and services. The figure is organised as a layered structure, with layers representing different levels of abstraction for a generic \ac{NFV} system (i.e., rather than the typical OSI layered network structure). As we move up from the Disaggregated Hardware towards the Application layer, each layer provides further levels of component abstraction.
Although many of the projects named in the figure are hosted by the Linux foundation, they originated independently, driven by different industry organisations, different business divisions and following different standardisation efforts and thus there can be substantial overlap in functionality across them. We can also observe that while most of the projects operate on single layers, three of them, \ac{CORD}, \ac{ONAP} and \ac{OPNFV} operate across multiple ones. While an in-depth overview of each project is outside the scope of this paper, we provide a brief description of some of the most popular ones across the layers.
\begin{itemize}
\item \textit{The \acf{TIP} \cite{tip_2018}}: operates at the disaggregated hardware layer, aiming at opening up the transmission system and separating its hardware and software components. While this idea resembles much that of OpenFlow, \ac{TIP} goes far beyond the layer2/layer3 switch, aggregating multiple sub-projects each addressing a different technology, covering wireless and wired transmission and from access to backhaul and core networks. For example, the Open Optical \& Packet Transport sub-project provides an \ac{OLS} specification for validating interoperability of optical transponders across multiple vendors.

\item \textit{The \acf{DPDK} \cite{intel_dpdk}}: is a set of user space libraries that can be used to speed up packet processing operations in general purpose processors. This is achieved through a number of optimisations, such as bypassing the Linux kernel and using technology like \ac{DMA} and polling of devices to avoid processing interrupts. Recent \ac{DPDK} performance reports \cite{DPDK_performance_2018} show its ability to process over 70 millions packet per second on a server with high-end Intel Xeon processor and Intel Ethernet network adapters.

\item \textit{The ONOS controller \cite{ONOS}}: is a network controller driven and supported by the \ac{ONF} and developed specifically for network and service providers. Some of its distinctive characteristics are design for high availability and resiliency, and high scalability to support several millions request at the northbound interface and low latency response time for network events. Its southbound interfaces are extensive and include most recent protocols such as OpenFlow, \ac{NETCONF}, P4, RESTCONF, while supporting well established ones, such as \ac{BGP}, \ac{SNMP}, \ac{TL1} and \ac{CLI}. Perhaps, one of the most interesting features recently implemented is the intent-oriented framework, which allows users to specify their control plane request in terms of policy rather than specific actions. This provides ONOS with flexibility in selecting the most adequate actions to meet the requirements of a given request (e.g., a connection of X Gb/s capacity and Y ms maximum latency  between two end points) and to respond appropriately should a network failure or congestion occur.
Finally, ONOS was the network controller of choice for the \ac{CORD} project, described further below.


\item \textit{The Openstack \cite{Openstack}}: is a cloud computing platform for the virtualisation of computing resources, typically used in data centres or smaller clusters. The system is made up of several components, each managing a different aspect or service of the cloud. For example, the "Neutron" system manages the networking aspect, "Nova" manages the computing resources, "Cinder" the storage system, "Keystone" provides user authentication services, etc. OpenStack enables integration of hardware components form multiple vendors and is arguably the most common open source cloud computing system to date. 

\item \textit{The \acf{OSM} \cite{MANO}}: is an Open Source implementation of the ETSI \ac{NFV} "Management \& Orchestration" reference architecture \cite{etsi_mano}. Its architecture is divided into three main functions, whose descriptions are briefly reported below, showing how services are mapped into sets of virtual functions and subsequently into physical hardware resources. 

The \textit{\acf{VNFM}} manages \ac{VNF} instances across their lifetime, operating tasks such as instantiation of \acp{VNF}, scaling them with respect to resource usage and terminating them when no longer needed.

The \textit{\acf{NFVO}} sits on top of the \ac{VNFM} and operates the task of selecting the correct VNFs and chaining them together to provide the service requested by a user or application. It also provides service monitoring, to assure the compliance with the given requirements. In \ac{OSM} the \ac{NFVO} and \ac{VNFM} are embedded in a module called \textit{Network Service Orchestrator}.

The \textit{\acf{VIM}} is a virtualisation framework whose main task is to keep an inventory of what physical resources (compute, storage and network) are assigned to the virtual network functions instantiated by the \ac{VNFM}. \ac{OSM} offers the possibility to use different \ac{VIM} implementations. The \ac{VIM} module developed by \ac{OSM} is the OpenVIM, which follows ETSI recommendations \cite{etsi_NFV_PER} and represents a minimalist and lightweight implementation of \ac{VIM} functionalities. OpenStack can also be used (i.e., instead of OpenVIM), offering additional flexibility and expandability, which however comes at the cost of a more heavyweight implementation (OpenStack was indeed designed  to handle large-scale compute, storage and networking resources). Plugins for other \ac{VIM} implementations are also supported, such as \ac{AWS} and VMware \ac{vCD}.
In terms of controllers support, the \ac{OSM} can operate with OpenDaylight, ONOS or Floodlight.

\end{itemize}

As previously mentioned, in Fig.  \ref{fig:Linux_fun_classification} we also find a number of projects spanning more than one layer, which typically make use of some of the components described above to provide a usable framework capable of implementing the cloud-CO concept. We describe below three main projects that have recently gained visibility across the telecommunications industry.

\begin{itemize}
\item \textit{The \acf{CORD} \cite{CORD}}: is arguably the first project that attempted to implement in software a virtualised central office. CORD used an implementation-oriented approach, in that it minimised the time spent  in defining architectural design and instead provided quickly after their launch an initial proof of concept, creating ferment among operators. Their success was endorsed by the fact that within few months from the lunch of their consortium, several major industry players joined their effort as partners or collaborators. As shown in Fig. \ref{fig:Linux_fun_classification}, \ac{CORD} covers both the virtual network management and the control plane (making use of ONOS for the latter). Their platform development is differentiated into three main application areas, providing use case specific implementations. The \ac{R-CORD} aims at providing broadband services and leads the development of the \ac{VOLTHA}, which virtualises the passive optical network. The \ac{M-CORD} provides solutions for the virtualisation of mobile networks, providing \ac{C-RAN} implementation with a number of different functional splits. The \ac{E-CORD} targets enterprise services, providing connectivity on demand to business customers (e.g., implementing \acp{VPN}, \ac{SD-WAN}, etc.).
While \ac{CORD} has to date differentiated across these three main areas, they are all based on the same overall framework and can reuse the same hardware infrastructure (e.g., commodity servers, storage and switches), with some differentiation on the I/O devices (i.e., a physical layer \acp{OLT} for \ac{R-CORD}, \acp{RRU} for \ac{M-CORD} or \acp{ROADM} and transponders for \ac{E-CORD}).

\item \textit{The \acf{OPNFV}}: is a carrier-grade open source reference platform for \ac{NFV}.  Similarly to \ac{CORD}, \ac{OPNFV} reuses open source components from the protocol stack in Fig. \ref{fig:Linux_fun_classification}. However, while \ac{CORD} is focused on pre-defined use cases and operates a specific selection of elements to provide production-ready solutions, \ac{OPNFV} has an open approach and relies on the platform end users for selecting and integrating the components to be used. In addition, \ac{OPNFV} is based on the OpenDaylight network controller and provides extensive support for the \ac{BGP}. With reference to Fig. \ref{fig:Linux_fun_classification}, we can see that \ac{OPNFV} covers all the layers of the virtualisation stack, confirming its main target of bringing together all \ac{NFV}-related activities into a coherent reference platform. In this perspective, \ac{CORD}'s focus on specific use cases has led its consortium to consider its participation in future releases of \ac{OPNFV}  (i.e., to produce CORD-based scenarios of \ac{OPNFV}).

\item \textit{The \acf{ONAP} \cite{ONAP}}: is a project developing an \ac{NFV} platform that was recently formed by the fusion of \ac{OPEN-O} led by China Mobile and \ac{ECOMP} led by AT\&T. \ac{ONAP} focuses on the management aspects, covering layers from the network data analytics down to the network control. It provides the ability to specify both orchestration and control frameworks (to automatically instantiate services) and an analytic framework for monitoring performance of the services created.
\end{itemize}

In order to put the three platforms above into perspective, we should notice that while they serve in principle the same high-level objective of Network Orchestration through NFV and SDN, \ac{OPNFV} may be seen as an architecture driven approach backed up by extensive test suites, while \ac{CORD} may be seen as a use-case driven approach. ONAP uses a top-down approach (building on the extensive \ac{ECOMP} functionality) with specific emphasis on enterprise requirements such as end-to-end and automated service activation.
We should also notice that the three platforms have all been developed independently and only recently brought within the Linux foundation umbrella. While their development carries on autonomously, their consortia have operated liaison efforts to provide future interoperability across the platforms. For example, as system integrator, \ac{OPNFV} could provide interoperability and end-to-end performance validation for the entire platform, while \ac{ONAP} could be adopted for the management and control aspect of \ac{NFV}.
Liaisons are also ongoing between \ac{CORD} and \ac{OPNFV}, that could see \ac{CORD}-based scenarios, focused on the use of merchant silicon and white-box switches, being included in future \ac{OPNFV} releases.




\vspace{3mm}

Among the benefits that the combined use of \ac{NFV} and \ac{SDN} has brought to the telecommunications world, one that stands out is the ability to provide convergence across multiple network domains in an unprecedented fashion.
A representative use case is the integration of fixed and mobile networks, where for example the scheduler of a \ac{PON} and that of a mobile \ac{BBU} can be synchronised to reduce the transmission latency over the \ac{PON} in a cloud RAN implementation \cite{NTT_FMC,Effenberg_FMC}. 
Another example \cite{vr_fronthaul} is the dynamic adaptation of the PON assured rate and mobile fronthaul rate, according to the actual mobile cell load. 
Several other examples exist, spanning from the convergence of mobile, optical and cloud compute resources \cite{Tzanakaki}, to multi-tenancy application of the \ac{cloud-CO} \cite{Bruno_OFT}, \cite{Vilalta}. 

Recently, this network convergence and virtualisation trend has progressed further down the stack, to the optical transmission layer, with the aim of opening up the transmission systems, including transponders, \acp{ROADM}, amplifiers, etc. This concept, called optical network disaggregation, is addressed in the next section, which discusses how the cloud-CO revolution is affecting the optical transmission layer.



\section{Optical Layer Control and Disaggregation}
Many future use cases for the development of novel connected applications, require the ability to provide instant capacity across the network. These vary from the provisioning of point-to-point bare capacity on demand to a business, to the establishment of an end-to-end \ac{VNF} chain from an end user to an edge or centralised computing centre; or to the migration of network functions and virtual machines across the access and metro area to fulfil capacity, latency and availability requirements of specific applications. 
In addition, the ability to provide dynamic capacity reconfiguration is essential to enable statistical multiplexing of networking and computing resources, which are typically constrained by economics.
Resource virtualisation platforms can quickly allocate capacity on demand, as far as the underlying physical layer has enough available resources. In data centres, such physical layer capacity is typically pre-provisioned, since in this environment ultra-dense and high-capacity connectivity is affordable. However, this is not the case for telecommunications networks, where fast dynamic reallocation of physical capacity will be a requirement to provide cost-effective, high-capacity and low latency connectivity, across multiple domains, to a cluster of new applications that will bring new revenue into the network \cite{Dohler_Latency,Ovum_latency}. 

The multi-domain challenge was partly addressed by work proposing \ac{SDN} orchestration of capacity across domains. The concept has been demonstrated with respect to both inter-domain control plane \cite{DISCUS-IDEALIST} and data plane across different network technologies (i.e., involving optical packet switching and flexgrid) \cite{PDP_strauss}. However, the optical layer was still considered a closed system, a black box with at best a proprietary network controller exposing northbound interfaces to the network orchestrator. 

More recently, work in \cite{Kiper_Li_PDP} has demonstrated the possibility to use southbound interfaces (e.g., OpenFlow) to directly control \ac{ROADM} nodes and gather information from \ac{OSNR} monitors to trigger re-routing at the control plane after physical link failure. This gives the controller the ability to carry out data analytics and the flexibility to directly control the physical transmission network. It can be seen as a step forward towards a completely open and disaggregated solution, where all optical sub-components are controlled through an \ac{SDN} control plane \cite{Kilper_Li_OFC}. 

The advantages of optical layer disaggregation are manifold and fit well the requirements of future network applications and services. From a cost perspective, optical disaggregation allows to source components separately, which in turns allows to avoid vendor lock-in and increases competition, which can drive down prices and improve component performance. It also carries advantages from a network convergence perspective, as it enables integration of systems developed from different vendors and for different network domains (i.e., access, metro and long haul).
From a network performance perspective, a greater control over the optical transmission network can be used to reduce end-to-end latency, for example by enabling the orchestration of transparent wavelength connections between access, metro and data centre domains. The network orchestrator could for example provide ultra-low latency paths from a location in the network access to a data centre by controlling the optical systems in both domains, to dynamically provisioning a transparent path that is terminated directly in the server that carries out the data processing for the service \cite{convergence_tutorial} (this scenario is further discussed in the next section). A typical example is the dynamic allocation of \ac{C-RAN} streams across edge and centralised computing \cite{Tzanakaki_xhaul}, following the latest functional decomposition architectures from the \ac{NGMN} \cite{NGMN_func_dec}. 

\begin{figure}[t]
   \centering
    \includegraphics[width=\columnwidth]{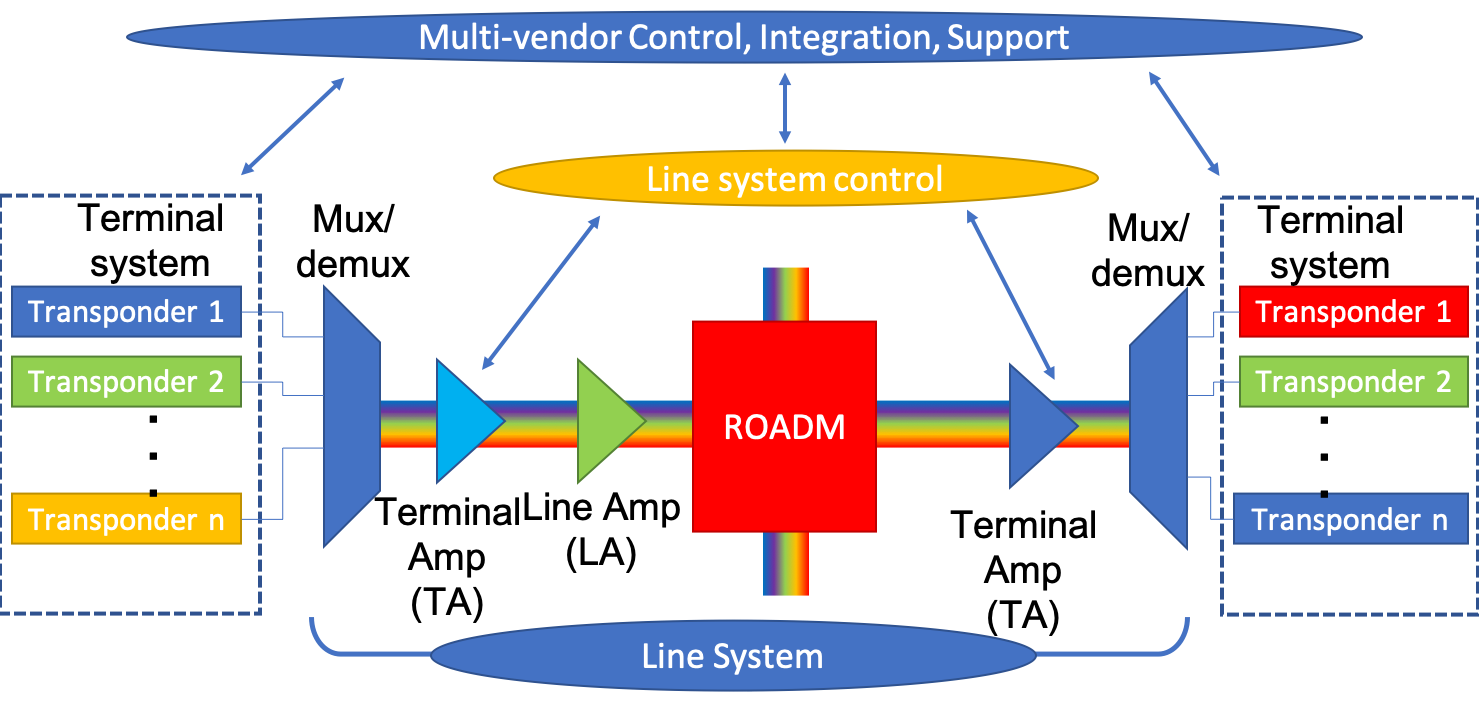}
    \caption{Illustration of the principle of full optical disaggregation}
    \label{fig:disag}
\end{figure}

Fig. \ref{fig:disag} illustrates the concept of full optical network disaggregation, where different network elements and control systems can be supplied by different vendors. While this might be by some considered the ultimate goal of an "Open Roadmap" \cite{Bennet}, current developments are taking up intermediate steps towards this goal. The \ac{OLS} for example assumes that the line system (inclusive of the line system control with reference to Fig. \ref{fig:disag}) is provided by one vendor, but transponders can be sourced by different suppliers. The aim of \ac{OLS} is thus to guarantee interoperability between any transponders and line systems.
A further step should see the opening of \acp{API} to provide control to external \ac{SDN}  systems. The OpenROADM \cite{Open_ROADM} specification, driven by a consortium of \ac{ROADM} vendors, envisages the use of \ac{NETCONF} with YANG data models to provide the controller with an abstract representation of the devices. In addition, the OpenROADM specification aims at disaggregating the \ac{OLS}, providing abstraction for functions carried out by \acp{ROADM}, transponders, pluggable optics, in line amplifiers and muxponders. 

The \ac{ONF}-based \ac{ODTN} \cite{ODTN}, mostly driven by operators, also aims at providing open \acp{API} and \ac{OLS} disaggregation (although its first phase only addressed point-to-point transmissions systems), but puts a stronger emphasis on the use of largely adopted interfaces, such as the \ac{TAPI} as northbound and OpenConfig-based models as southbound interfaces. In addition, \ac{ODTN} uses ONOS as reference controller. (Additional information on network models in support for optical disaggregation can be found in \cite{Szyrkowiec}).

While there are benefits to the disaggregation of the optical layer, we notice that its implementation has lagged behind with respect to the rest of the network. The main reason is arguably that there is a fundamental difference between virtualising layers operating in the digital domain (e.g., above L2) and the optical transmission layer that operates in the analogue domain and needs to address optical transmission impairments.
Thus, before the optical layer can be fully opened and its control system integrated with the rest of the layers, there are challenges to be solved. For example, in a closed optical communications system the same vendor provides transponders, in line amplifiers and \ac{ROADM} nodes, thus the variability and unpredictability of the system is minimised. This is important especially for longer links (e.g., in the long-haul) where the available optical margins are squeezed to a minimum. Thus, when we attempt to open up the optical systems, making use of components from different vendors, typically the variability and uncertainty of performance of an end-to-end path increases, reducing optical margins and making the network less efficient.
As of today there are ongoing discussions on the feasibility and benefits of disaggregating optical networks, and most recognise the existence of a trade-off between the need for increasing the amount of transmission monitoring components to reduce the system uncertainty, and the cost they add to the network. Studies in \cite{OFC_Morris} provide some evidence that the benefits of optical layer disaggregation are more likely to occur in the metro area, where the optical margins are less strict than in the regional and core networks.
In addition, much work was recently carried out on the use of machine learning techniques to improve prediction of \ac{QoT}, so that the expected performance of a new path can be assessed before it is provisioned (\cite{Tomkos_ML,Pointurier_ML,Tornatore_ML,Panayiotou_ML}, only to cite a few; the reader can also refer to \cite{Musumeci_ML_tut}, \cite{Velasco_ML_tut} for comprehensive tutorials on this topic). Overall, while this is a promising area of research, more work is required on assessing how to collect, store and share adequate data sets to train the machine learning algorithms.

\vspace{3mm}

In conclusion, work on the disaggregation of optical networks is at an early stage but ongoing, with several consortia involved in the definition of interfaces or interoperability specification.




\section{Sample use cases}

If we look at the progress carried out in the cloud networking area (e.g., at the network virtualisation for data centre domains), we see mature products (e.g., the VMware network virtualisation and security platform - NSX \cite{NSX}) being deployed, allowing a complete abstraction of the network, thus enabling full portability of the sub-networks across racks and data centres, improved resilience, enhanced security, etc. Moving down the stack, we see research oriented to the virtualisation of physical transmission devices for telecommunications networks, aiming to provide high granularity, isolation and quality differentiation of slices sharing the same physical channel, with examples both in the wireless \cite{slice_radio} and optical \cite{Ou_JLT} domains.

Virtualisation and slicing will play a fundamental role in future generation of networks, allowing network customisation down to the flow levels to provide personalised network services to specific classes of applications and to enable true multi-tenancy. However, their full performance can only be delivered if the underlying physical transmission channel is capable of quickly adapting paths and capacity to the requirements of the layers above. This is especially true in the access and metro parts of the network, where the large number of end-points does not allow to pre-provision all necessary capacity between them. Incidentally, these are also the parts where optical layer disaggregation seems more practical \cite{OFC_Morris}.
While work is ongoing on the definition of frameworks, architectures and interfaces \cite{Open_ROADM,tip_2018,ODTN}, more research is needed in dynamic provisioning of QoS-oriented (e.g., with bounded latency and availability levels) physical layers links across different technological domains. These include the mobile access, the fixed optical access, the optical metro transport and the computing domains (both at the edge and in large centralised \acp{DC}). 

In this section we provide technical details of two specific use cases of resource slicing in next generation \acp{cloud-CO}. 
\subsection{\ac{PON} scheduling virtualisation for multi-service and multi-tenant applications: the \ac{vDBA}}

The first use case focuses on the virtualisation of a specific component of a cloud-CO architecture, the \ac{DBA} running in the \ac{OLT} network element. 
As mentioned above, \ac{R-CORD} was the first project to propose and implement a virtualised \ac{OLT} as part of their central office virtualisation framework. However, in their implementation the \ac{DBA} for upstream scheduling of capacity is implemented in the hardware and runs as a single instance. In \cite{OFC_vDBA}, our research group at \ac{TCD} has introduced the concept of \ac{DBA} virtualisation, and proposed and developed a testbed implementation architecture in \cite{Slyne_OFC}. \ac{vDBA} provides a mechanism to move scheduling algorithms from a single instance running in hardware into multiple instances running in software. 
This allows different \acp{VNO} sharing the same physical PON infrastructure to fully control, down to the intra-frame microsecond time scale, the capacity scheduling of their access network slice.
For example, addressing a possible 5G use case, a \ac{VNO} could run two different scheduling algorithms, one optimised for customers requiring low latency operations and another optimised for efficient use of resources.
The \ac{vDBA} concept was ratified in the \ac{BBF} standard TR-402 \textit{"Functional Model for PON Abstraction Interface"} \cite{vDBA_BBF}.

Our testbed was implemented by linking a server running the \ac{vDBA} mechanism in software to an \ac{OLT} running on \ac{FPGA} hardware, implementing XGS-PON framing and 10Gb/s optical transmission. 
The \ac{vDBA} allows more than one \ac{DBA} algorithm to run in parallel, each producing its own virtual \ac{BMap}. These are forwarded to a second function, called \ac{ME} which merges them together producing a physical \ac{BMap} that is forwarded to all \acp{ONU} over the physical layer.

\begin{figure}[t]
   \centering
    \includegraphics[width=\columnwidth]{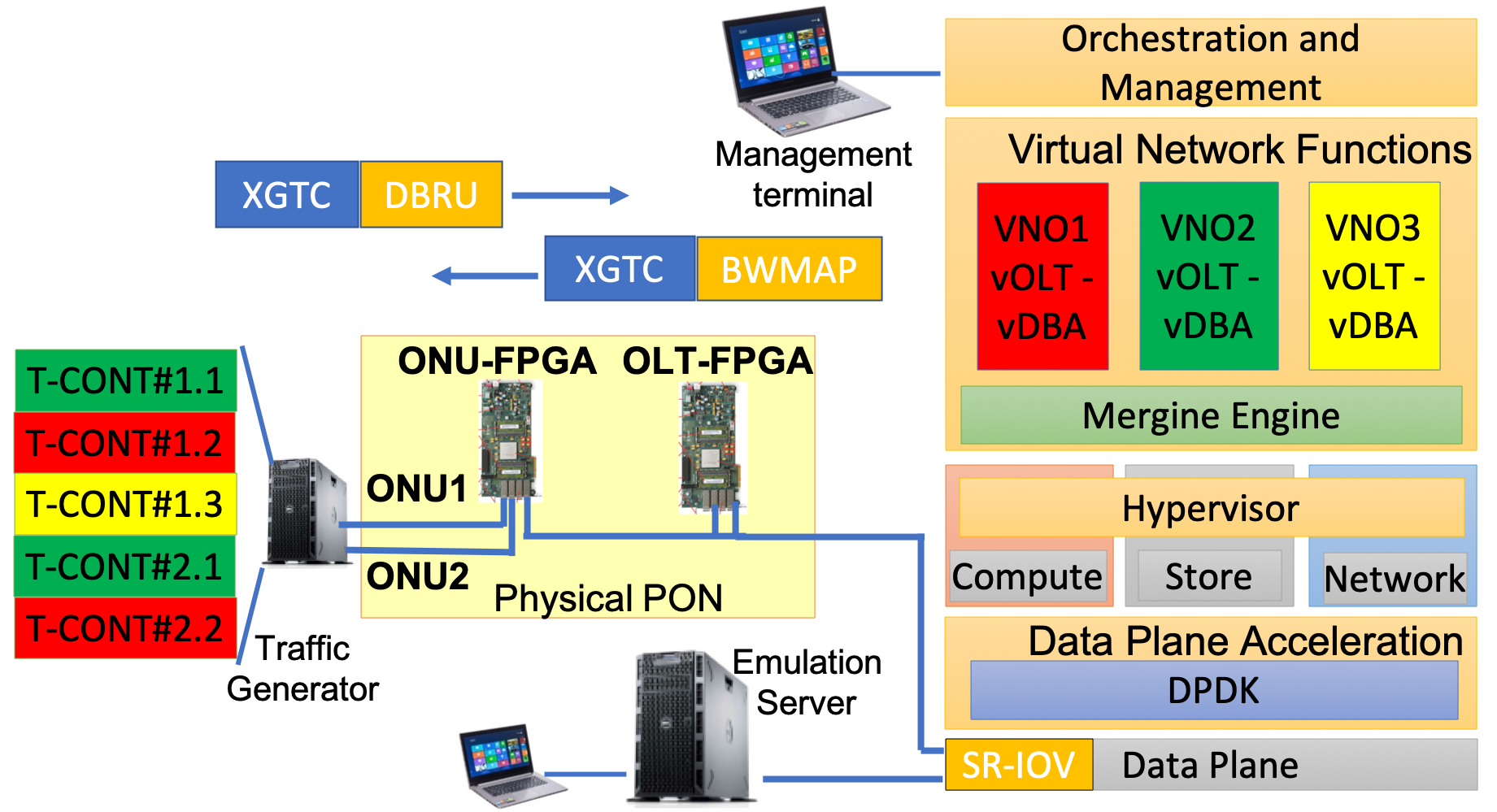}
    \caption{Experimental vDBA demonstrator developed at \ac{TCD}}
    \label{fig:vdbatestbed}
\end{figure}

Our implementation, shown in Fig. \ref{fig:vdbatestbed} uses OpenStack as  a virtualisation framework on top of which the \ac{vDBA} and \ac{ME} functions run. OpenStack provides extensive options on the choice of hypervisor and virtualisation technologies. Because the \ac{vDBA} and \ac{ME} functions are compute intensive and do not require much disk storage, we use lightweight virtualisation based on \acp{LXC}. 

One of the main issues with off-loading the \ac{DBA} to an upstream host is that, even for a low to medium sized PON tree, the hardware side of the OLT injects high levels of small datagram packets into the host's network packet processing module (e.g., in the tens or hundreds thousand per second). Network cards and Linux network kernel modules can cater for large traffic streams of several gigabits per second when transmitted over large-sized packets. However, the small packet characteristic of the \ac{DBRU} (potentially sent every frame by all \acp{ONU} to the \ac{OLT}) can consume a large amount of \ac{GPP} resources, because each incoming \ac{DBRU} packet generates a hardware interrupt (which represents one of the main causes of delay in software packet processing systems).

Our implementation thus optimised the \ac{vDBA} and \ac{ME} modules for network packet processing \cite{vDBA_ECOC}.  Firstly, we use \ac{SRI-OV} to off-load packet processing interrupts to the network card, and thereby minimise interrupts to the CPU main processor and cores. We make extensive use of the \ac{DPDK} software libraries, developed by Intel, to reduce the amount of unnecessary copying of data between memory on the same host (even if the memory segments are associated with different Virtual Machines) as well as to optimise locking in the reading from and writing to buffers and queues. We prevent real-time critical functions such as the Merging Engine from being continuously interrupted, by using \ac{DPDK} to assign individual or complementary functions to distinct cores. 

\begin{figure}[t]
   \centering
    \includegraphics[width=\columnwidth]{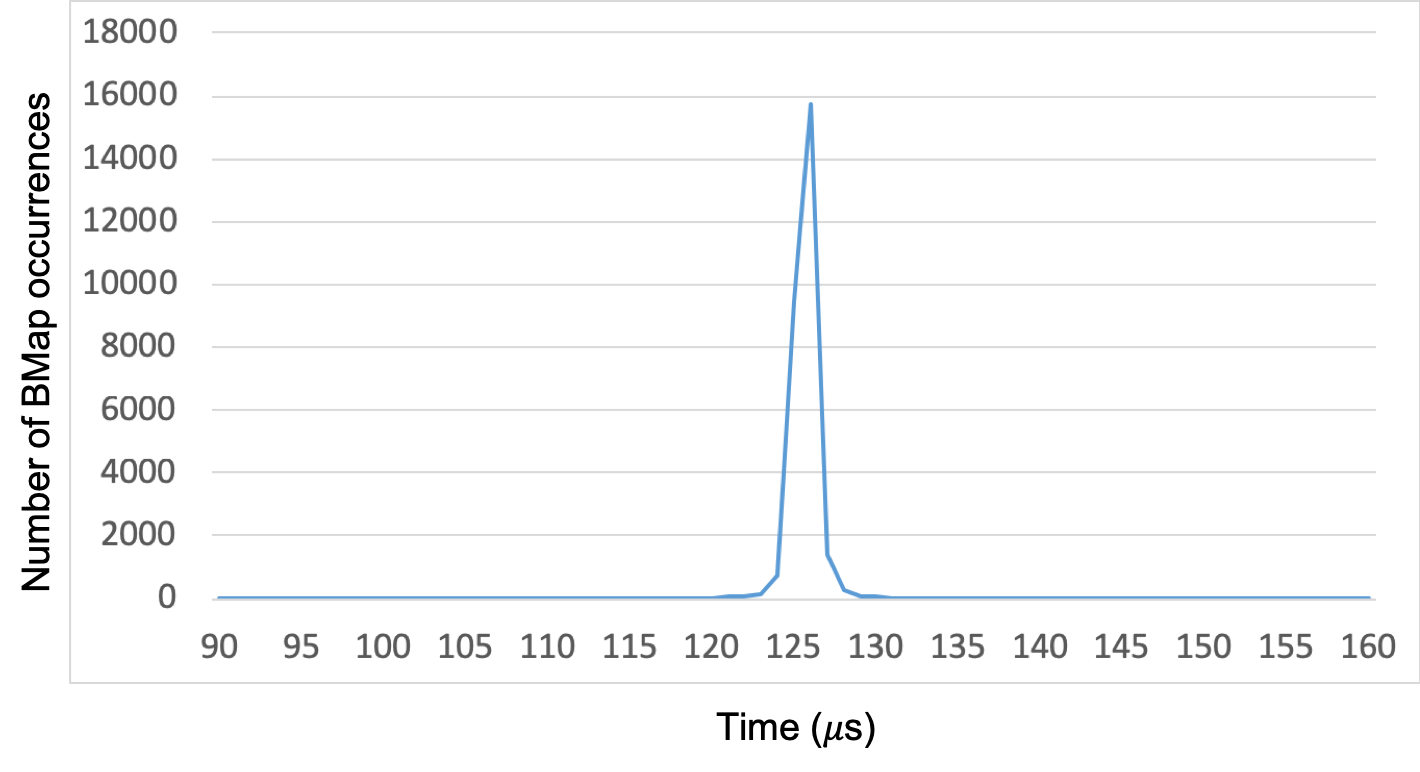}
    \caption{Bandwidth Map stability characterisation}
    \label{fig:interbwmap}
\end{figure}

To demonstrate the functionality and the performance of our \ac{vDBA} architecture, we generated a constant stream of \ac{DBRU} traffic based on a combination of real traffic at the ONU and a traffic emulator, to reproduce the scenario of a \ac{PON} with 32 \acp{ONU}. Fig. \ref{fig:interbwmap} shows the interval, in microseconds, between the transmission of successive \acp{BMap}, calculated over 30,000 bandwidth generation cycles. The average inter-transmission time takes into account the packet routing within the host platform between the Merging Engine and vDBA applications, as well as the processing time for the calculation of the constituent Bandwidth Maps by 2 \acp{VNO}, and the subsequent merging into a single bandwidth map by the Merging Engine. As it can be seen in the figure, we obtain minimal variation in the \ac{BMap} transmission time (with a calculated variance value of 0.75). This shows that the \ac{vDBA} \ac{BMap} generation is highly stable, and moving from hardware to software does not deteriorate the PON performance. 

\subsection{Cross-domain optical operations in dynamic access/metro/cloud environments}
\begin{figure}[t]
\begin{subfigure}{.5\textwidth}
   \centering
        \includegraphics[width=\columnwidth]{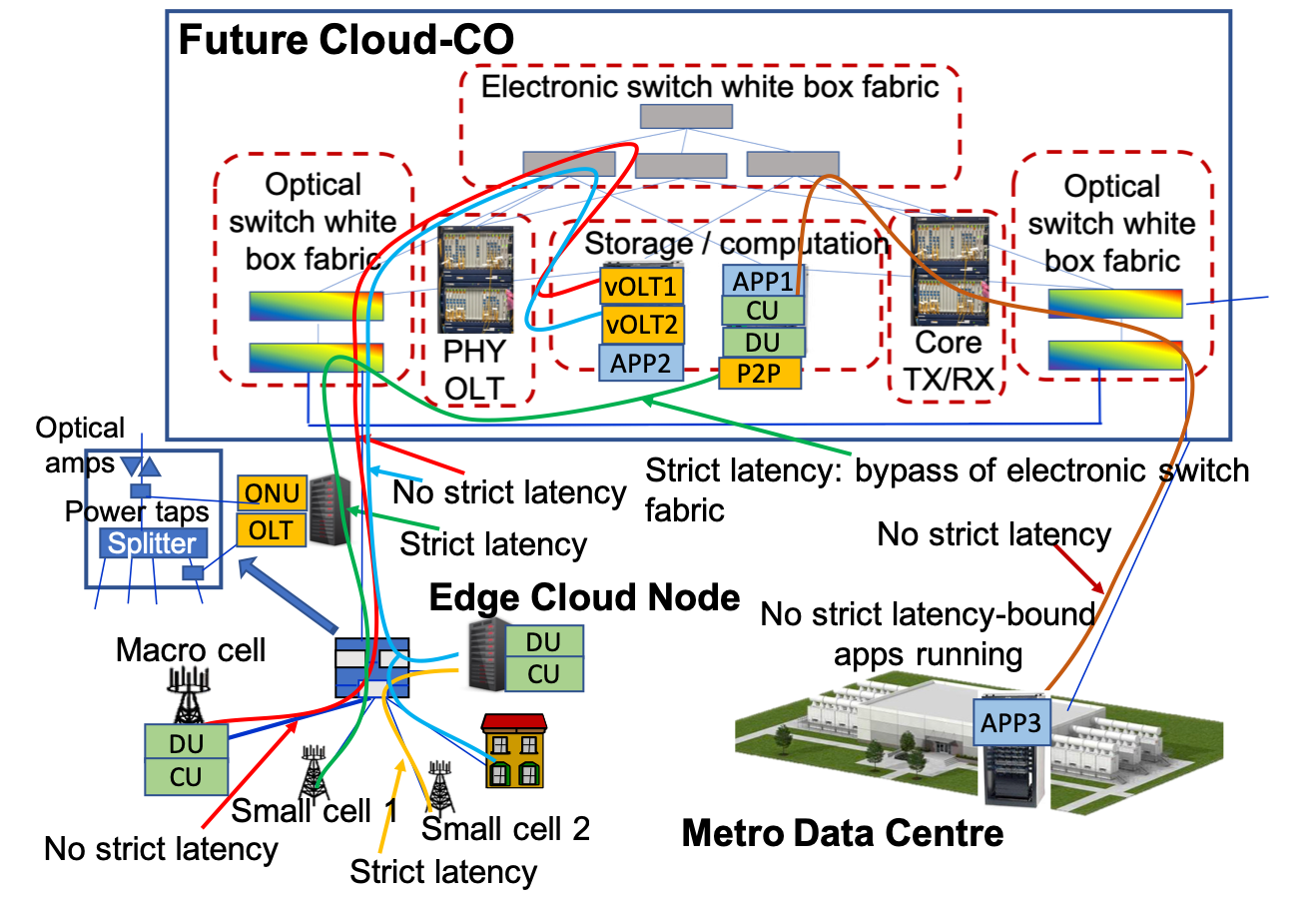}
    \caption{}
    \label{fig:scen1}
    \end{subfigure}
    
\vspace{3mm}

\begin{subfigure}{.5\textwidth}
   \centering
        \includegraphics[width=\columnwidth]{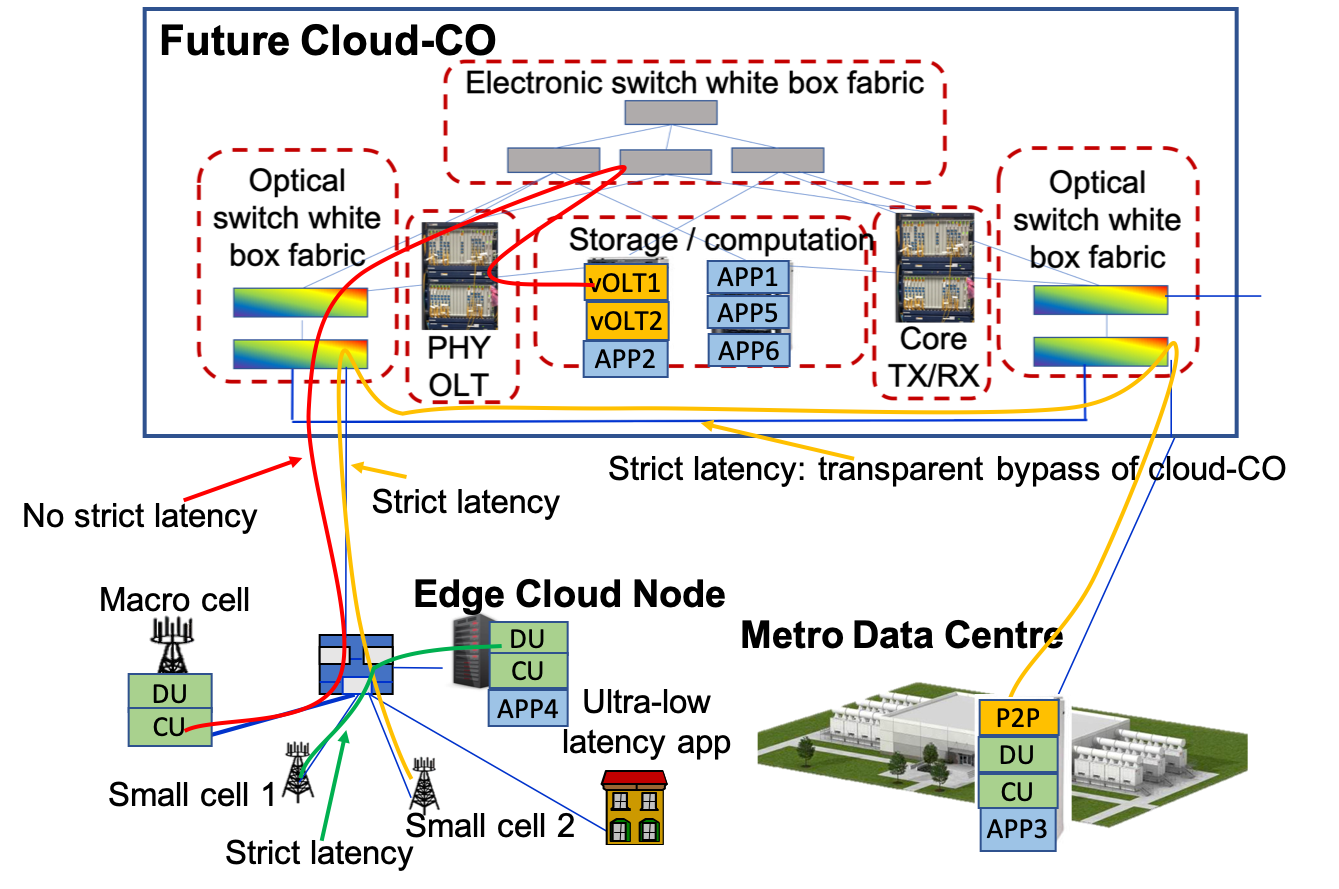}
    \caption{}
    \label{fig:scen2}
    \end{subfigure}
    \caption{Future \ac{cloud-CO} with dynamic optical layer scenario}
    \label{fig:hypothetical_future_scenario}
\end{figure}

The second use case focuses on an end-to-end dynamic slicing scenario, from mobile access to edge or central cloud. 
For this we consider a future \ac{cloud-CO} with dynamic optical switching capabilities and high densification of mobile cells, as shown in Fig. \ref{fig:hypothetical_future_scenario}. Following the functional decomposition guidelines \cite{NGMN_func_dec}, different functional split options are available where \acp{DU} and \acp{CU} can be dynamically placed, depending on compute resources available and latency constraints. The cells are backhauled through a virtualised \ac{PON}, which can provide links to both local computation elements (e.g., the Edge Cloud Node) and the \ac{cloud-CO}, which is linked to the rest of the metro network and can provide connectivity to larger metro \acp{DC}. 

The scenario reported in Fig. \ref{fig:scen1} shows a macro cell with co-located \ac{DU} and \ac{CU}, which does not require strict latency connection (shown in red colour); thus it is terminated at a physical \ac{OLT} in the \ac{cloud-CO} and from there it reaches its destination server through the shared electronic switch fabric (e.g., vOLT1 and then APP2 in the figure). Small cell 2 operates on a split number 7 \cite{NGMN_func_dec} (i.e., the split option that divides the physical layer into two, the low PHY and the high PHY), which requires a low latency connection (shown in yellow) to the \ac{DU}/\ac{CU}, which runs in the Edge Cloud Node. It should be noticed that the PON splitter allows for optical signals to be tapped locally, so that specific wavelength channels can be terminated at a local Edge Cloud Node. From there, the data stream continues towards the cloud-CO with no strict latency constraints (blue line) and it is thus mixed with residential traffic and terminated at an \ac{OLT} (e.g., vOLT2 in the figure). For small cell 1 instead, as there is no spare capacity available at the edge node, a transparent connection through the PON is required (shown in green). Since in this instance the \ac{cloud-CO} cannot provide low latency connectivity through its electronic switch fabric, the link is operated as \ac{P2P} directly to the server where the \ac{BBU} is located (e.g., the blocks labelled \ac{P2P}, \ac{DU}, and \ac{CU}, respectively, in the figure). The detailed architecture of the optical switch fabric is not shown in the figure, but examples are available in works such as \cite{DISCUS_JOCN} addressing access/metro convergence and \cite{Heilios, Philip_Ji} looking at hybrid electrical/optical switching in data centres.
Also, in this use case no strict latency connection is required towards the metro data centre, which runs high-layer applications. A simple example of a PON splitter node implementation is also shown in the figure, with power taps that can link the edge node both to end-user access points and to the CO. 
More complex and flexible implementations using active wavelength steering (i.e., \acp{WSS} and \acp{ROADM}) could also be considered in some parts of the \ac{ODN} as their cost decreases over time.

Fig. \ref{fig:scen2} shows the same use case on a different time, where a high-priority application (APP4 in the figure) with low latency requirements needs to run for a user connected to small cell 1. The control plane needs thus to preempts small cell 2's connection to free space to run \ac{DU}, \ac{CU} and application locally at the Edge Cloud Node for cell 1. Since at this time the cloud-CO cannot provide computational resources capable of guaranteeing the low latency processing at the DU for cell 2, the cloud-CO operates a full optical bypass \cite{IP_bypass}, which provides a point-to-point transparent link connection (shown in yellow) of small cell 2 directly to a server in the metro \ac{DC} \footnote{Due to the latency constraints associated to a functional split of the physical layer, it is typically considered that the propagation distance from small cell to data centre might not exceed 40 km. However, the transmission latency constitutes only a small part of the latency budget, and this value could be increased if the optical bypass can reduce latency of other operations. For example, we should consider that an optical transparent link will remove the latency associated with electronic switching in the DC. In addition, the ability to provide more powerful processing resources in the metro DC would decrease the overall processing latency, leaving a higher latency margin for the optical link transmission.}. 
This terminates on a \ac{P2P} transceiver directly in the server (or rack) where the BBU functions and applications are processed (labelled \ac{P2P}, \ac{DU}, \ac{CU} and APP3, respectively, in the figure).

\section{The business case for telecomms digitisation}
As it can be expected, the main driver towards virtualisation of the operators' networks is the prospective increase in network cost-effectiveness and revenue streams. While there is today still uncertainty on which aspects of the central office virtualisation will provide the largest economic advantage, network strategists have identified some of the main features that will drive  cost reduction  and  revenue increase across the entire telecommunications digitisation process.
The \ac{WEF} Digital Transformation Initiative has estimated that the biggest cost savings, in terms of network infrastructure and energy consumption, will be generated by moving operations from dedicated hardware resources to software running on commodity servers. Their estimate predicts an overall contribution to profit over the next 10 years of \$200bn \cite{WEF_WP}. 
In addition, they predict that network automation will reduce many of the operational expenses, generating an extra cumulative profit of \$75bn. 
Additional cost savings are expected from enhanced security, which will reduce the costs related to data breaches, generating estimated profits of \$80bn.
In addition to cost savings, the network digitisation will provide new revenues streams. \textit{Research \& Markets} estimates that opening up the network \acp{API} will allow third parties to provide improved services, generating annual revenues of more than \$200bn by year 2022 \cite{RM_report}. In addition, in \cite{Creaner_book} the author estimates that new services to residential and enterprise customers, including future \ac{IoT} applications, will generate global profits of the order of \$300bn over the next decade.

Another means by which the digitisation of the telco industry will radically enhance the economy is through the improvement of \ac{IT} services across all types of businesses. A study carried out by the Harvard Business School \cite{Harvard}, showed that other non-telco industries can reap substantial benefit by leveraging new data analytics to improve customer relationships, internal operations, product creation and delivery, and management of human resources. 

In conclusion, the medium-to-long term expectations are indeed of substantial economic benefit across several industry sectors. However, it is to be expected that in the short term the transition between current models and next generation virtualised services will bring about some inefficiencies, due to duplication of network functions and expertise across the old and new platforms.

\section{Conclusions}
This paper, extending the work in \cite{ECOC_tutorial}, has provided a description of recent trends on the virtualisation of telecommunications networks. After a brief introduction of the historical background that has led over the past two decades to the development of the \ac{SDN} and \ac{NFV} concepts, the paper has described ongoing work around the idea of cloud central office, providing a classification of the main frameworks and platforms currently under development. It has then provided a link to the still largely unexplored domain of optical network disaggregation, which can be considered a natural evolution towards a fully open network stack. 

In section IV we have provided two examples of use cases for network virtualisation. The first, based on experimental testbed results, showed the possibility to use virtualisation to disaggregate a PON network so that multiple tenants and multiple services (e.g., from residential to \ac{C-RAN}) can be accommodated over a shared infrastructure. The second showed the principle of using disaggregated optical transmission to cut across optical domain boundaries allowing for highly dynamic network reconfiguration to help meet the high capacity and low latency requirements of next generation services. We can also foresee similar scenarios in the future occurring for several users and applications, all with different priorities and requirements but competing for the same shared resources: the network will need to accommodate them as they are powered on and off arbitrarily. 

Thus, depending on the time of day and other variables, both network functions and applications might need to be re-routed across the network, while trying to maintain their capacity and latency constraints. A full orchestration across wireless, optical and computing domains is thus required to maximise the number of services that can run successfully in the network. From an optical level perspective, this requires an agile layer capable of creating transparent connections across several \acp{ROADM}, capable of crossing \acp{DC} boundaries to provide minimum latency towards a given processing unit and capable of using the minimum optical bandwidth that is necessary to provide the link (i.e., using the most suitable modulation format and flexgrid bandwidth allocation). 

Additional research is also required to provide bounded latency services for \ac{NFV} flows within a cloud-CO. The trend of moving into software the services that were previously implemented in hardware does bring issue of latency bounds within a functional chain. 
If a \ac{cloud-CO} based on common data centre architectures needs to disaggregate and virtualise hardware components, it will need to deliver bounded network and processing performance for some of the \acp{VNF}. In addition, upcoming applications linked to virtual and augmented reality will only exacerbate such low latency requirements. 
While some recent work has focused on \ac{NFV} orchestration across domains and \ac{VNF} placement optimisation \cite{NFV_placement}, the research community still needs to develop frameworks to scale QoS tools to cope with several million flows, within an automated framework that spans multiple network domains.




Finally, as the \ac{cloud-CO} becomes more and more integrated with access and metro optical networks, we envisage that dynamic, potentially disaggregated, optical networking will become part of the solution, providing low-latency links for high-capacity 5G and beyond mobile access over a multi-service, multi-tenant, statistically multiplexed network infrastructure.

\section*{Acknowledgments}
This publication has emanated from research conducted with the financial support of Science Foundation Ireland (SFI) under Grant Numbers 14/IA/2527 (O'SHARE) and 13/RC/2077 (CONNECT).

\ifCLASSOPTIONcaptionsoff
  \newpage
\fi

\end{document}